# Symmetrical-geometry constructions defining helicoidal biostructures. The case of alpha- helix.

## Mikhail Samoylovich[a] and Alexander Talis[b]

[a] Central Scientific Research Institute of Technology "Technomash", Moscow
[b] A.N.Nesmeyanov Institute of Organoelement Compounds of Russian Academy of Sciences, Moscow
Correspondence email: miksamoylovich@gmail.com

**Abstract**
The chain of algebraic geometry constructions permits to transfer from the minimal surface with zero instability index, and from the lattice over the ring of cyclotomic integers to the tetra-block helix. The tetra-block is the 7-vertex joining of four tetrahedra sharing common faces; it is considered as a building unit for structures approximated by the chains of regular tetrahedra. The minimality condition of the 7 – vertex tetrablock as a building unit is the consequence of its unique mapping by the Klein's quartic (which is characterized by the minimal hyperbolic Schwartz triangle) into the minimal finite projective geometry. The topological stability of this helix provided by the pitch to radius ratio H/R of $2\pi/\tau^2$ ($\tau$ is the golden section) and by the local rotation axis order of 40/11 = 40 exp(-H/R). These parameters determine the helix of $C_\alpha$ atoms inside the $\alpha$ – helix with the accuracy of up to 2%. They explain also the bonding relationship $i \rightarrow i+4$ between the i-th amide group and the (i+4)-th carbonil group of the residues in the peptide chain and the observed value of the average segment length of the $\alpha$-helix which is equal to 11 residues. The tetra-block helix with the N, $C_\alpha$, C′, O, H atoms in the symmetrically selected positions, determines the structure of the $\alpha$ – helix. The proposed approach can display adequately the symmetry of the helicoidal biopolymers.

1. **Introduction**

Steric interactions of molecules related to their sizes and forms impose rigid structural restrictions upon the ways of positioning in the *3D* Euclidean space $E^3$. Such interactions determine to a large extent the packings of molecules into helices widely distributed in the biologic world [1-4]. An important role is played also by steric interactions of molecules, related to their sizes and shapes, and imposing strong structural limitations on space positions of molecules. Moreover, these interactions to a large extent determine the packing of molecules into helices, which are widespread in biologic objects [1, 2, 5].

Among helicoidal biological structures stand apart due to their extraordinary stability the DNA and the α-helix which is stabilized by hydrogen bonds between the *i-th* peptide amide and the *(i+4)-th* groups of the residues. For the ratio of pitch of the helix H to the radius R equal to ≈2.35, a non-crystallographic axis 18/5 suggested by Pauling with the angle of helical rotation of $100^0$ is realized in [6]. Similar to a crystal determined by lattice parameters and a set of crystallographic axes, the α-helix is determined by: 1) the ratio *H/R* of helical pitch to radius, 2) the axis of helical rotation m/p, 3) the bonding ratio *i → i+4*, 4) the observed average segment length of the *α*-helix of 11 residues, 5) certain positioning of atoms in the peptide plane [1].

In terms of Van der Waals radii the α-helix is partially approximated by the helix of tube of radius *ρ* with pitch *H* and radius *R*. The ratio of such a helix to the volume of a cylinder of radius R determines the packing density of the tube helix, which reaches its maximal value 0.784 for *ρ ≈ R* and the pitch angle $\theta_{max}$ determined by the relation *H/2πR* and equal to *18.1⁰*. At the same time, inside the helix a central empty channel formed of radius *0.025R*. Of the helical biopolymers considered within such an approximation the most densely packed is the α-helix with packing density 0.781 [7, 8]. Thus, upon approximation of the α-helix by a helix of polyhedra of radius ρ, the following conditions must hold: 6) of close radii $\rho \approx R$ and 7) the existence of small sized empty central channel. Proteins can be considered as dense packings of more or less similarly sized sphere-like units – amino acids, approximated within a polyhedral representation by a packing of tetrahedra. The densest packing of 600 regular tetrahedra is achieved in a *4D* polyhedron (polytope) *{3,3,5}*. The connection of its substructures with the α-helix has been demonstrated in [3, 4].

A necessary condition for existence of a crystalline structure is its ability to be embedded in $E^3$, which is ensured by the existence of the space group – the group of discrete motions of the space $E^3$ preserving distances between points. Similarly, a necessary condition for existence of the α-helix must be existence of mathematical constructions determined by properties of $E^3$ and defining a (topologically stable) helix virtually independent of the sequence of amino acids in the polypeptide chain. As a confirmation of this supposition may serve the data of comparative analysis of mean squared displacements of atoms as well as the B-factor reflecting the extent to which the density reconstructed around it is wider than in an ideal model. They show that both flexible and rigid segments of *3D* protein structure are conserved in the process of evolution. Theoretical calculations of the B-actor as well as the mean squared displacements have shown also that they are mostly determined by folding (of polypeptide chain into a spatial structure) and protein structure and virtually do not depend on the sequence of amino acids in a polypeptide chain [9]. In the general case, the criterion of dense packing of spheres or tube-like helices does not determine the secondary structure of protein, to which the α-helix also belongs [10]. However, the summary symmetry basis of the approaches mentioned has allowed to assume that the structure of the α-helix must be conditioned by some helical structure determined by a special surface in $E^3$ as well as a discrete *4D* structure.

Studies of the α-helix and other helical biopolymers have answered some questions concerning their structure, but the problem of adequate symmetry-based explanation of extraordinary stability of such systems is still important. In work [11], on the structural level was mapped to a system of constructions of algebraic geometry and topology, allowing for construction of a topologically stable helical packing of 7-vertex unions or regular tetrahedra determining the basic parameters of the $\alpha$ - helix. Constructing also the models of A, B and Z - DNA within this approach has shown a common symmetry basis for such helical biopolymers. Realization of such an approach that determines on a symmetry level the stability of linear biopolymers has required the deepening and extension of the formalism developed in [11].

The structure of a crystal is defined by a structural realization of one of the space groups – a mathematical construction not dependent on the existence of atoms and their interactions. The present paper is devoted to the definition of the α-helix as a structural realization of a non-crystallographic symmetry construction.

## 2. MINIMAL SURFACE WITH ZERO INSTABILITY INDEX COMMON TO BOTH HELICOID AND CATENOID

It has been shown in [11], that in order to describe helical structures it is necessary to use the most general minimal ruled surfaces, the helicoid and the catenoid whose embedding in $E^3$ defines polytope. Upon projection of $S^3$ into $E^3$ the vertices of a polytope are positioned on the surface of the catenoid – the minimal surface of zero mean curvature. All non-congruent complete minimal ruled surfaces form one-parameter families of helicoids for which the pitch *H* of a helix of radius *R* can be chosen. Upon decrease of *H* to a certain value $H_{cr}$ there will be a moment when the film ceases to be a helicoid and spans an additional surface between helices. Thus there exists a unique minimal surface common to both the helicoid and the catenoid, which is characterized by certain (critical) ratio of the pitch of the helix to its radius [11, 12].

In order to describe all minimal surfaces spanning the circuit consisting of two coaxial circles of radius $R$ positioned in parallel planes separated by the distance $H$, it is sufficient to describe all its spanning catenoids. For small $H$ there is a point of bifurcation $H_{cr}$ in the system, and it determines two configurations: the stable one that is close to a cylinder and an unstable one that is close to a cone. This point is given by the unique positive solution of the equation [12]

$$\text{cth}(H_{кр}/2R)=(H_{кр}/2R)= \pi/\tau^2 = k \approx 1.2 \text{ (more precisely } 1.1996786\ldots) \text{ ,} \qquad (1)$$

linking 2 constants: $\pi$ and golden section $\tau=(1+\sqrt{5})/2 \approx 1,618$. Thus, for a certain critical value $H_{cr}$ a unified configuration is realized – a film spanning each of the circles by a plane disk [12]:

$$H_{кр}/R= 2\pi/\tau^2 \approx 2,4 \quad \text{and} \quad H_{кр}/2\pi R = \text{tg } \theta_{кр.} \approx 0.38187 \approx 1/\tau^2 \quad , \qquad (2)$$

where the pitch angle $\theta_{cr.} \approx 20,906^0$ is the angle between a 2-fold and a 3-fold axis of the icosahedron. Thus, such a film is the only minimal surface common to both the helicoid and the catenoid and is determined by the special pitch angle $\theta_{cr} \approx 20,906^0$ of its defining helices. The conditions (1), (2) define a special point of bifurcation nature. It determines the condition of transition from a locally minimal to a locally cylindrical surface, for which neighborhoods of every point are approximated by a cylinder-like surface. To describe all minimal surfaces tightening (spanning) the contour of two coaxial circles of radius $R$ positioned in two parallel planes a parted by a distance of $H$ it is sufficient to define all catenoids spanning this contour. With decreasing interturn spacing at certain value $H=H_{cr}$ the bifurcation point is arising. Making a significant simplification, we shall assume that the relation (2) determines the transition from a locally minimal to a locally cylindrical surface, namely, the surface, for which the neighbourhood of every point is approximated by a cylindrical surface. Using the points belonging to both the catenoid and the helicoid is an algebraic approach allowing for the use of conjugate surfaces and the introduction of finite (discrete) constructions.

The stability of a minimal surface is characterized by its surface instability index (*Ind M*) corresponding to the number of topologically distinct ways of changing its area. A minimal surface *M* is stable if any continuous variation on its boundary increases the area of *M*. Instability of *M* increases with the growing *Ind M*, equal to 1 for the catenoid and ∞ for the helicoid. For minimal surface *M* its stability is determined by the possibility to change its area by small strains. Stability of *M* is characterized by the index *Ind M*, which correlates to the number of ways to change the surface area. If this index is not zero, the surface *M* is unstable. The instability of the *M* surface increases as the *Ind M* increases which is equal 1 to a catenoid and to ∞ for a helix. There are well developed methods to construct complete minimal surfaces, embedded in $E^3$, by using Weierstrass' representations [12]. Combined with introduction of an exterior metrics (which is physically equivalent to fixed distances between atoms, molecules or their centers) the stability of *M* also ensures the existence of certain type of stability also for the volume bounded by *M*. The surface *M* common to both the catenoid and the helicoid is a complete minimal surface and can be constructed using Weierstrass representations [12]. It is shown [12] that Weierstrass representations allow one to define catenoid as well as complete helicoid, and, in general case, an associated family for some minimal surface *M* consists of locally isometric minimal surfaces (incongruent pairwise, as a rule).At certain conditions one can create a configuration as joining of helicoid and catenoid.

Assume *M* is determined by a global Weierstrass representation, defined among other parameters by a domain *U* of the complex plane *C*. According to [12], if for a minimal surface $M_0$ with a finite instability index the image of a domain *U* under a Gaussian map is contained in a certain spherical belt *Q* of the sphere $S^2$ (of unit radius), then *Ind $M_0$=0*. Let *M* be some surface given by Weierstrass representation and $U \subset C$ is some subdomain of the complex plane. The surface *M* is characterized by *Ind M=0* if the image of the *U* in the some open submanifold of the $S^2$ sphere [12]. For the condition *Ind $M_0$=0* to hold it is necessary also that the surface $M_0$ be compact (namely, without an edge) or compact but with an non-empty boundary and not closed. The spherical belt *Q* must be contained between two parallel planes removed from the center of the sphere by the distance th(k) and cutting from it two neighborhoods of its poles, each of about 1/12 of its area. In this case the spherical belt *Q(k)* of the sphere $S^2$ constitutes 5/6 of its area. That submanifold can be defined as submanifold onto a part of the $S^2$ sphere (about of 5/6 of the total sphere area) confined between two parallel planes. Indeed, such planes are aparted from the sphere center over the distance th$t_0$, where $t_0$ – is the unique root of the equation cth$t_0$=$t_0$, and cut off the domain of the about 1/6 of the sphere surface area.

A change from *H* to $H_{cr}$ determines a set of coaxial catenoids embedded in the projection $S^3$ into $E^3$. The sphere $S^3$ is topologically equivalent to the groups *SU(2)* (of complex matrices 2x2), whose principal bundle space is the sphere $S^7$ embedded into $E^8$. Summing up the above, the relations are obtained that define the topological basis of the desired construction

$$CP^1 \rightarrow C \supset U \rightarrow Q(k) \subset S^2 \subset S^3 \leftarrow S^7, \qquad (3)$$

where $CP^1$ is the complex projective plane diffeomorphic to the sphere $S^2$, corresponding to the projective completion of the complex plane *C* by a point at infinity. Arrow denotes (homomorphic) maps into the subset in question.

For a helicoid *M* the Cartesian coordinates of points are expressed via hyperbolic functions, which in the end determine the necessity of using fractional linear transformations. In particular, while preserving the conformity of the mapping, $S^2$ may be viewed as a complex projective line, for which the mapping $S^2 \rightarrow S^2$ is a Mobius transformation. It is given by a fractional linear transformation *f(z)*, equivalent to the product of translation by *d/c*, inversion, rotation with dilation (dilation or compression), and translation by *a/c*,

$$f(z)=(az+b)/(cz+d)=(z+d/c)\,(1/z)\,((bc-ad)z/c^2)\,(z+a/c), \qquad (4)$$

where *z* is a complex variable; *a,b,c,d* are complex numbers *(ad-bc≠0)* and $f(z)^{-1}$=*dz-b/-cz+a*. Upon imposition of the restriction *ad-bc=1* all Mobius transformations correspond to the map of an open disk *(|z|<1)*. Each finite subgroup of the group of fractional linear transformations corresponds to polyhedral (point) groups in $E^3$. In the dimension *2*, the orientation-preserving Mobius transformations are exactly the maps of the Riemann sphere. Additivity and invariance for hyperbolic motions are possible for angles with a shared vertex that can be expressed via the area of the corresponding sector of rotation multiplied by the metric coefficient *k=1.2*. [13]

A surface is called isothermal if isothermal coordinates can be determined in the neighborhoods of all its points (except for some singular points). For the biologic structures in question the term "isothermal" is equivalent to the term "equipotential". It is assumed that no potential difference arises between sections of a stable *3D* structure. Such structures relate to positioning putative packing centers on isothermal nets, which correspond to Villarceau circles on torus [3, 4]. A conformal diffeomorphism $S^2 \rightarrow S^2$ is a Mobius transformation corresponding also to projective transformations of $RP^3$ if isotherms are considered. Minimal surfaces are isothermal and triangulated isothermal surfaces are invariant with regard to Mobius geometry [14]. Hence it is sufficient to limit one's attention to such triangulated isothermal minimal surfaces, using for its description the Mobius geometry in its quaternion implementation: putting into correspondence with quaternion degeneration a pair of reciprocally inverse structures, and an inversion of $S^n$ with an involution (*n≥2*). Furthermore, all periodic minimal surfaces have an infinite index of instability [12], hence it is necessary to use constructions of locally periodic (locally lattice-like, to be more precise) systems with finite total curvature. This also dictates the necessity to apply discrete differential geometry, in particular, its Mobius interpretation [15].

# 3. LATTICE OVER THE RING OF CYCLOTOMIC INTEGERS, SUBSTRUCTURES OF THE POLYTOPE {3, 4, 3} AND THE PARAMETRIC AXIS 40/11

A surface $M_0$ defines an ordered *3D* helical packing given that the singular points of $M_0$ are related by symmetry transformations of corresponding vector lattices satisfying (1)-(3). For symmetries that relate singular points of $M_0$ it is sufficient to consider symmetries determined by substructures of the *8D* lattice of octonions $E_8$ closing the series of possible numbers: real – complex – quaternions – octonions [16]. For an n-dimensional lattice the coordination sphere $S^{n-1}$ defines a system of vectors, and, consequently, an n-dimensional polyhedron – a regular or non-regular polytope [17]. The symmetry groups of *4D* polytopes are symmetry groups of fractional linear transformations [18]. The polytope *{3,4,3}* is closest to the extremum of the volume functional corresponding to the *4D* ball, hence the polytope {3,4,3} should be chosen as a basic polytope for a discrete implementation of the construction of $M_0$ [11,16,17].

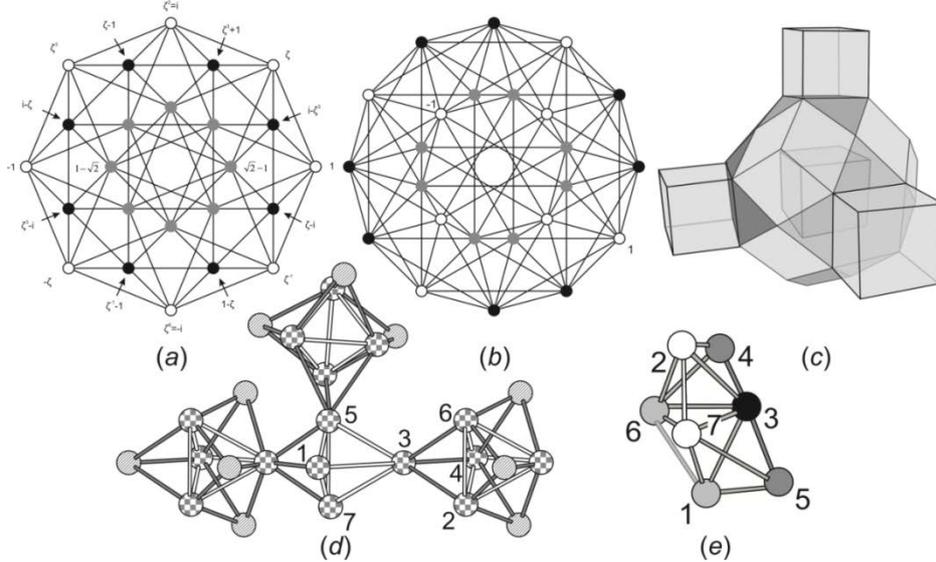

*Figure 1.* a) *Projection onto the complex plane of the polytope {3, 4, 3}, whose vertices are represented as elements of a Gaussian lattice. Interior and exterior octuples of vertices are vertices of the polytope {4, 3, 3}, the middle octuple of vertices are vertices of the polytope {3,3,4} (Adapted from Figure 8.1 [16])*
*b) Projection onto the complex plane of the polytope {3, 4, 3} whose vertices are redistributed into two 12-vertex chains. Discarding the vertices corresponding to 1 and -1 leads to formation of two 11-vertex chains.*
*c) A cube in the polytope r{3,4,3} is surrounded by 8 cubes bordering it by vertices and 6 cuboctahedra bordering it by faces. Their centers lie in the sections of the polytope {3, 4, 3}, starting from a vertex: a point, the North cube, and the equatorial octahedron. The polytope r{3,4,3} consists of 24 cubes and 24 cuboctahedra filling voids between cubes. 3 cubes are selected standing on faces of 3 cuboctahedra bordering the central cube. The centers of the 11 cubes (except the central one), are the vertices of the chain shown in Figure b.*
*d) A chain of tetrahedral stars from the polytope sn-{3, 4, 3}, containing a chain of "internal" regular tetrahedra (shown as squared balls), where each tetrahedron borders the preceding one by a vertex and the succeeding one by a face. Exterior vertices of tetrahedral stars are dashed balls. The vertex 3 is common to the tetrahedra 3246 and 3157 forming a united 7-vertex object.*
*e) In the 7-vertex union of 4 regular tetrahedra – a tetra-block $\Delta_1$ between the tetrahedra 3246 and 3157 (shown in bold lines), the tetrahedra 1456 and 1567 are positioned.*

One of the locally periodic lattices satisfying the requirements listed above is a lattice of the form $Z[\zeta]$, embedded in C over the ring of cyclotomic integers, where $\zeta=exp\pi i/4$, $\zeta^2=i$ and $\zeta^4=-1$. It is a variant of the real lattice $D_4$ and upon mapping $1 \to 1100$, $\zeta \to 0110$, $\zeta^3 \to$ -1001 the vertices of the projection of the polytope *{3, 4 ,3}* onto the complex plane may be represented by 24 elements from *Z[ζ]* (Figure 1a), which can be identified with 24 minimal vectors $D_4$ of norm 2. The factor manifold $D_4/3D_4$ consists of 81 cosets: a zero coset; 24 cosets represented by unique vectors of norm 2; 24 cosets represented by unique vectors of norm 4; and 32 cosets each of which contains exactly 3 vectors of norm 6. There is a one-to-one mapping of the elements of the ring *Z[ζ]/3Z[ζ]* onto the elements of the field $F_9 \times F_9$: elements of norm 2 onto elements of norm – 1; elements of norm 4 onto elements of norm 1, and elements of norm 6 (3 times) onto nonzero elements of norm 0. Thus, $D_4 \leftrightarrow Z[\zeta] \to Z[\zeta]/3Z[\zeta] \to F_9 \times F_9 \to F_3$, where the field $F_9$ consists of the elements 0 and 8 powers of *ζ* [16].

Elements of norm 6 from *Z[ζ]/3Z[ζ]* are put in correspondence with the 96 vertices of a semiregular polytope – rectified *{3,4,3}* (*r{3,4,3}*) consisting of 24 cubes and 24 cuboctahedra. Mapping of the 96 vertices of *r{3,4,3}* into 32 elements of $F_9 \times F_9$ corresponds to a distribution of 96 vertices into 32 triangular faces isolated from each other. The centers of the cubes are the vertices of the starting polytope *{3, 4, 3}*, and the centers of the cuboctahedra are vertices of the dual polytope *{3,4,3}*<sup>*</sup>. Cubes border each other only at vertices, cuboctahedra only at triangular faces, and cubes with cuboctahedra on square faces (Figure 1c]). Upon sectioning, beginning from a vertex, of the polytope *{3,4,3}* by the plane $E^3$, one obtains: the vertex, corresponding to the north pole of $S^3$, 8 vertices of the north cube, 6 vertices of the equatorial octahedron, 8 vertices of the south cube and the vertex corresponding to the south pole of $S^3$ [17]. The vertices of the polytope {3, 4, 3} correspond to the 24 elements of *Z[ζ]*, where ±1 corresponds to the polar vertices, and *Z[ζ]\±1* to the remaining 22=2×11 vertices of the north (south) cube and the equatorial octahedron. Each vertex of *{3,4,3}* is the center of a cube in r{3,4,3}, hence the removal of the two polar vertices in *{3,4,3}* leads to removal of the two polar cubes (i.e., of 16 vertices denoted by ±8) in *r{3,4,3}*. Discarding 16 of 96 vertices corresponds to discarding 1/6 of the sphere, allowing one to map the remaining 80 vertices *(r{3,4,3}\±8)* onto the spherical belt *Q (k)*. Then *Q (k)* will contain two 40-vertex manifolds corresponding to the north and south unions of 11 cubes of *r{3,4,3}*. It can be demonstrated that the manifold *r{3,4,3}\±8* possesses symmetries of the form (4). Thus, concretizing (3) for the polytope *{3, 4, 3}*, the relations are obtained

$$RP^2 \subset CP^1 \supset U = Z[\zeta]\backslash\pm 1 \subset Z[\zeta] \to Z[\zeta]/3Z[\zeta] \to (r\{3,4,3\}\backslash\pm 8) \leftarrow Q(k) \subset S^2 \to RP^2, \quad (5)$$

where *Z[ζ]\±1* is the union of the two 11-vertex figures from the polytope *{3,4,3}*. Such an 11-vertex figure *({4} ∪ {3} ∪ {4}')* is formed by the squares *{4}* and *{4}'* (two edges of the polar cube subsuming all its vertices) and the triangle *{3}* (the north face of the equatorial octahedron) while interpreting the polytope *{3,4,3}* as a union of its sections starting from a point. The sequence of such sections is as follows: the north pole, a cube, the equatorial octahedron, a cube, the south pole [17]. The square *{4}* corresponds, for instance, to the set *{4}<sub>ζ</sub>* of elements $\zeta^{2n}$, *n=1, 2, 3, 4*, the triangle *{3}* to the set *{3}<sub>ζ</sub>* of elements *ζ, ζ(1-ζ), ζ(1-√2 )* of the lattice *Z[ζ]*. In this case, the 11-vertex figure *(Z[ζ]\±1)/2* is the union

$$(Z[\zeta]\backslash\pm 1)/2 = \Delta_\zeta \cup \varphi_\zeta \Delta_\zeta, \quad (6)$$

of the two 7-vertex figures $\Delta_\zeta \leftrightarrow (\{4\} \cup \{3\})$ intersecting at the shared triangular face: $\Delta_\zeta \cap \varphi_\zeta \Delta_\zeta = \{3\}_\zeta$, with $\varphi_\zeta$ corresponding to an element from *Z[ζ]*, which can be given a representation in the form *exp2πikp/m*.

The Gosset construction transforms each cuboctahedron of the polytope *r{3,4,3}* into an icosahedron, and a cube into a tetrahedral star – the central tetrahedron on each of whose faces stands another tetrahedron [17]. An external vertex of one star is a vertex of the central tetrahedron of another, hence all 2×48 vertices of the polytope *sn-{3,4,3}* thus obtained lie in two 12-star unions of intersecting stars or 2x12 central tetrahedra of these stars, isolated from each other. Discarding the 8 vertices of the polar star from the union leads to a partitioning of the remaining 40 vertices among 7 isolated tetrahedra and 4 isolated triangles, i.e. tetrahedra with one vertex removed. The vertices not belonging to the central tetrahedron of the vertex of the polar star correspond to vertices of the cube taking every other vertex; hence isolated triangles that appear when such a vertex is discarded (which corresponds to defining such a vertex as stable under the given transformations of the point) cannot be adjacent. Thus, for every 7 and 4 elements *(Z[ζ]\±1)/2* there are respectively a quartet and a triple (not adjacent to other triples) of vertices of the polytope *sn-{3,4,3}*.

The vertices from *(sn-{3,4,3}\±8)/2* can be related to tetrahedra and to triangles isolated from one another. In two adjacent tetrahedra the succeeding one may be partitioned into a vertex and a triangle that, when added to the first tetrahedron, gives a unified 7-vertex figure. Similarly, the 7-vertex figure Δ' corresponding to $\Delta_\zeta$ is formed upon addition of an isolated triangle to a tetrahedron. Viewing Δ' as a union of two tetrahedra with common vertex, we obtain a homogeneous chain of tetrahedra where each tetrahedron borders the preceding one at a vertex and the succeeding one by a face. Therefore, *(sn-{3,4,3}\±8)/2* may be mapped into a chain constructed of 7-vertex figures *Δ'* sharing triangular faces (Figure 1d).

The tetrahedron is the simplex of $E^3$, and the union of tetrahedra by faces corresponds to a simplicial complex [19]. A possibility of putting a simplicial complex in correspondence to the surface $M_0$ under consideration is determined by the fact that in the bifurcation point given by (1) the topological regularity is broken and a cell structure forms on the corresponding manifold [19]. Centering the 24 icosahedra of the polytope *sn-{3,4,3}* leads to the polytope *{3,3,5}* – a partition of $S^3$ into 600 regular tetrahedra, which allows for a transition from the 7-vertex figure Δ' to a 7-vertex union by vertex of 2 tetrahedra from *{3,3,5}*. When joining in it the nearest vertices of the two starting tetrahedra by edges, two more tetrahedra appear, and the union of tetrahedra by face corresponding to the simplicial complex is formed. We shall call such 7-vertex union by faces of 4 regular tetrahedra (Figure 1.e) a **tetra-block** and denote it $\Delta_l$. The common vertex of the 4 tetrahedra will be termed the center of the tetra-block.

Generation of the helix by a tetra-block means that every succeeding tetra-block is attached by its face to the preceding one according to a single law. It can be demonstrated that *(sn-{3,4,3}\±8)/2* will correspond to a 40-vertex «U-helix» of 11 tetra-blocks, and *sn-{3,4,3}\±8* will correspond to the union of two such *U-helix*. In (3) a two-valued map of elements of $S^3$ onto $S^2$ is used, hence to lift the degeneracy when constructing a surface in $E^3$, the union of two *U-helices* must be doubled. Correspondingly, *(sn-{3,4,3}\±8)/2* may be constructed a 80-vertex manifold *(tr{3,4,3}\2·(±8)/2* of the semiregular 192-vertex polytope *tr{3,4,3})*, as well as a 160-vertex manifold *tr{3,4,3}\2·(±8)*, where each of 40-element vector sets correspond to a *U-helix* of 11 tetra-blocks. Each such helix borders the preceding and the succeeding one, therefore the first and the last tetra-blocks of the *U-helix* common with adjacent *U-helices*.

For the locally lattice-like (locally periodic) systems, elements of an abelian group modulo $2\pi i p$, generated by a rotation by the angle $p(360^0/m)$, are given by $\{exp2\pi ikp/m \mid k=0,1,2...m, exp2\pi ip=1\}$. The parameter *m/p* determines a local, conditional axis *m/p* giving the angle of helical rotation $p(360^0/m)$ with simultaneous shift along the axis by *1/m*. One of the conditions of the finiteness of the instability index of a minimal surface is that the corresponding polynomials (giving the surface elements) are fractional rational, which is ensured by representing *m/p* as a periodic decimal fraction *[m/p],(adc...)*, where *[m/p]* is the integral part of the fraction [12]. For instance, 40/11=3,(63), 30/7=4,(285714). For a crystal the period is a translation, which is possible only for a whole number of elements *[m/p],(0)* on a turn.

The local axis of the helix *m/p* is one of axes of Gosset helicoids:

$$m/p=2^\gamma \, 8I_n/4k_{js}m_{js} = 2^{\gamma+1} \, I_n/k_{js}m_{js}, \qquad (7)$$

where $2^\gamma 8I_n$ and $8I_n$ are numbers of vertices from the 2$^{nd}$ and 1$^{st}$ coordination spheres of the lattice $E_8$; $\gamma=0,1,2$; $I_n$, $I_s=k_{js}(m_{js}+1)$ are invariants of $E_8$, $k_{js}$ is an integer, $m_{js}$ are indices of sublattices embedded in $E_8$ [20]. In particular, for $\gamma=0$, $I_n=30$, $k_{js} m_{js} =2 \cdot 11$ and $k_{js} m_{js} =4 \cdot 5$ we obtain $m/p=30/11$ and $m/p=30/10=3$; for $\gamma=1$, $k_{js}m_{js}=14$ we obtain $m/p=30/7$. For $\gamma=1$, $I_n=20$, $k_{js}m_{js}=2\times 11$ we obtain *40/11*.

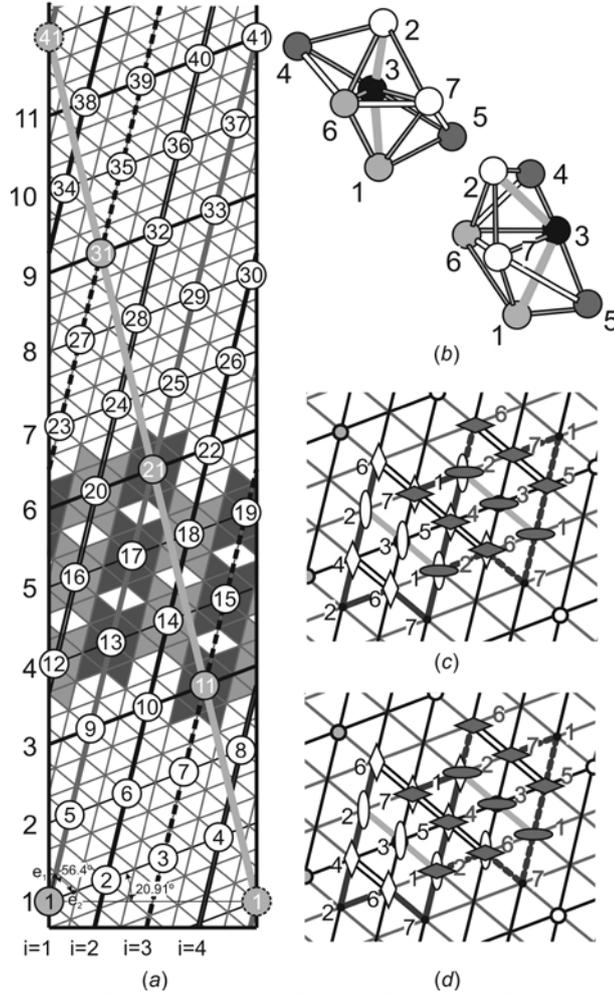

**Figure 2**. *a) Development of the triangulated cylindrical surface containing the 40/11 helix – a homogeneous distribution of 40 vertices into 11 turns. The turns shown in bold type contain 3 vertices each and the remaining ones four vertices each. The vertices i and i + 4 belong (taking into account identification of the vertical boundaries of the band) one of the four dashed i-lines, i = 1, 2, 3, 4. Intersections of turns with the solid gray (diagonal) line partition the helix into four 11-vertex parts having one common vertex. Vertices 11-21, making up of such parts are the centers of the tetra-blocks shown in gray and dark gray and forming a U-band. Adjacent gray and dark gray developments of tetra-blocks are connected by a 2-fold axis. The vectors $e_1$ and $e_2$ are unit vectors, and the length of the difference vector $e_1 - e_2$ equals 0.93.*

*b) Two tetra-blocks whose developments are embedded into the development a) are united by their shared face into an 11-vertex figure. In each tetra-block the chains of white and gray edges of length 0.93 form helices of the 15/4 type. The remaining edges are of unit length.*

*c) Development of the union by face of two tetra-blocks. The type of the chain 15/4 is preserved: chains of the white rhombi and ellipses of the 1$^{st}$ tetra-block are united, respectively, with the chains of the gray rhombi and ellipses of the 2$^{nd}$ tetra-block.*

*d) Development of the union by face of two tetra-blocks changing the type of the chain 15/4: chains of the white rhombi and ellipses of the 1$^{st}$ tetra-block are united, respectively, with the chains of the gray ellipses and rhombi of the 2$^{nd}$ tetra-block. Numbers of vertices in Figs. b-d coincide.*

For a triangulated isothermal minimal surface given by the function $F(x,y)$ its conformal parametrization is given by the equality of the scalar products of partial derivatives $(\partial F/\partial x, \partial F/\partial x) = (\partial F/\partial y, \partial F/\partial y) = \exp(u)$ [15]. For the case in question $u=-2k$, which (up to 0.1%) determines the norm $1/11 = \exp(-2\pi/\tau^2)$ for $F(x,y)$ and, finally, homogeneous distribution of the domain of the isothermal minimal surface given by (5) into 11 turns. Note that the local axis *40/11* satisfies both (7) and the relations

$$m/p = 40/11 = |sn\text{-}\{3,4,3\}\backslash\pm 8|/|Z[\zeta]\backslash\pm 1| \approx 40\exp(-2\pi/\tau^2), \qquad (8)$$

but for arbitrary axes (7) the relations of type (8) may not exist.

The Gaussian mapping of the tetra-block $\Delta_1$ determines a non-regular triangulation of the sphere by 15 edges into 10 triangles such that 6 triangles meet in one point, and in 3 pairs of vertices – 5, 4 and 3, respectively (Figure 1e). According to [11], in the plain development of the triangulated cylinder surface containing the helix 40/11 each tetra-block corresponds to a union of 10 triangles whose center coincides with the center of the tetra-block. Developments of two tetra-blocks united by a face are mapped into each other by a 2-fold axis, and the development of the *U-helix* of 11 tetra-blocks determines a triangulated *U-band*. The union of turns of the 40/11 helix is determined by the analytical continuation of the corresponding regular function describing one turn along some curve partitioned into overlapping segments. In particular, in constructing a complete helicoid from separate turns the upper border of the preceding turn is glue to the lower border of the succeeding one.

For the case in question this condition is realized by embedding into a 160-vertex development of triangulated cylinder surface a union of 4 overlapping *U-bands*, each of which is the union of 11 developments of tetra-blocks. Two overlapping *U-bands* possess a common center of the development of a tetra-block, hence the centers of developments of tetra-blocks form the development of the 40/11 helix – a partitioning of 40=4x11-4 points into 11 turns. Positions of the turns satisfying all of the above conditions are shown in Figure 2a.

**4. The tetra-block is the mathematically determined building unit of the structures approximated with the tetrahedral chains**

The minimal regular 7-vertex partition, where a tetra-block is embedded, is a partition of the torus into 14 triangles or the map *{3, 6}$_{2,1}$*. The map *{3, 6}$_{2,1}$* is one of the two congruent maps constituting the map *{6,3}$_{2,1}$* – a 14-vertex partition of torus by 21 edges into 7 hexagons (Figure 3.a,b). This map represents the incidence graph of the minimal finite projective plane *PG(2,2)* with the automorphism group *PSL(2,7)*. This graph is generated at: (1) collating of points and lines belonging to PG(2,2) to white and black vertices; (2) connecting by edges only white-black pairs which have an incidence sign at the intersection of corresponding line and column if the incidence table in Figure 3c. In this fashion, columns and lines of the *PG(2,2)* incidence table correspond to white and black vertices of the incidence graph, and incidence signs correspond to the edges of this graph.

In the partition *{6, 3}$_{2,1}$* of the torus into 7 hexagons a handle can be selected representable as a curved trigonal prism that complements the sphere to a torus [21]. Removal of this handle consisting of 3 edges leads to disappearance of one hexagon from the remaining sphere and the formation of the i-th non-regular partition of the sphere into 6 hexagons (Figure 3.a) – the map *{6,3}$^{3(i)}_{2,1}$*. This map may be considered the incidence graph of a Euclidean sub-configuration of the minimal finite projective plane *PG(2,2)*. The vertices of the tiling *{6, 3}* partitioning the plane into hexagons belong to the compound tiling denoted by the symbol *{6,3}[2{3,6}]*. In this symbol *[2{3,6}]* implies that two tilings *{3,6}* (each of which partitions the plane into triangles) are taken with common center and their vertices form a tiling *{6,3}* [17]. Similarly, a compound

$$\{6,3\}_{2,1}^{3(i)}[2\{3,6\}^{6(i)}{}_{2,1}]= \{3,6\}^{6(i)}{}_{2,1} \cup \varphi_i\{3,6\}^{6(i)}{}_{2,1} \qquad i = 1,2, \ldots I. \quad \varphi_i^2=1, \qquad (9)$$

map represents a union of the "white" and "black" composite maps $\{3,6\}^{6(i)}{}_{2,1}$ defining "white" and "black" tetra-blocks. Thus, the bichromatic map $\{6,3\}_{2,1}^{3(i)}$, uniquely defined by the incidence table of *PG(2,2)* without the i-th set of 3 incidence signs of *(PG(2,2)$_{3(i)}$)* defines a compound tetra-block – the union of the "white" and "black" tetra-blocks (Figure 3a). The possibilities for different unions of tetrahedra appear only when attaching a 4[th] tetrahedron to the 3 already existing ones [22] and lead to formation of two enantiomorphic (right and left) linear tetra-blocks $\Delta_i$, *i=1,2* and a plane tetra-block $\Delta_3$ (Figure 4c).

In the triangulated polyhedron tetra-block signified as $\Delta_i$ two adjacent triangles form a "rhombus" with their common edge as the short diagonal of the rhombus. Flipping *f* of a given diagonal (keeping vertices, edges and triangular faces numbers) determines the substitute of a short diagonal with a long one, and inversely the long diagonal is substituted by the short one. That transformation is not a rigid movement keeping spacing between points constant, and represents a Möbius transformation (4) which is determined by *PSL(2,7)*. This operation *f* transforms the tetra-block into a triangulated polyhedron which in common case is not a tetra-block. Therefore, the structures consisting from such polyhedra could be addressed as approximated with tetrahedral chains.

In the present paper Weierstrass two-periodic functions for elliptical curves are used implicitly. These functions give the most complete description of corresponding surfaces in complex coordinates. The spacing between catenoid basements is determined in the singular bifurcation point (for a catenoid and helicoid which is locally diffeomorphic to catenoid), this spacing is taken as ab inter-turn spacing $H_{cr}$. Simultaneously this value (with scaling R by 1) can be considered as the metric factor *k* in the hyperbolic projective space, and as an invariant of conformal transformations. According to (1), $2\pi = 2k\tau^2$, therefore the conversion of $2k$ radians into grades determines the golden angle [24]: $2,4 \cdot 57,3^0 = 360^0/\tau^2 = 137,5^0$, ensuring both the conjugation of elements in the spiral turn, and the conjugation between turns. The said above means the determination of two parameters *k* and $\tau^2$ of the surface defined in complex isothermic coordinates. These two parameters assign both real and imaginary periods. The corresponding Weierstrsass function takes real values at a real argument value.

Among all similar elliptic curves of genus 3, the Klein quartic defined by the $x^3y+y^3z+z^3x=0$ equation possesses the maximal automorphism *PSL(2,7)* group. This equation defines using of the Klein quartic for the transfer from the addressed surface to the polyhedral structure presentation with the tetra-block packing. The *PSL(2,7)* group is the symmetry group of the regular tiling of the genus 3 (3 handles) sphere. This tiling has 56 vertices and 24 heptacycles, viz the *{7,3}$_8$* map (Figure 3d), defined by the Klein quartic. The *PSL(2,7)* group is homomorphic to the triangular *(2,3,7)* group which is defined by the minimal hyperbolic Schwarz triangle. The $\pi/2, \pi/3, \pi/7$ angles of this triangle are realizing the tiling of the hyperbolic plane with reflections by its edges (Figure 3d). This group is the non-Euclidian crystallographic group (the discrete subgroup of the group of hyperbolic transformations of a plane) with the triangle as the fundamental domain [23, 25]. According to [23, 26], any projective *PG(2,p)* plane have the bichromatic $\{6,3\}_{p,1}$ incidence graph, therefore it is mapped into the regular torus tiling into hexagons. Among *PG(2,p)* the *PG(2,2)* and *PG(2,8)* planes are only embedded (Figure 3e) into the Klein quartic by the supermapping. The *PG(2,8)* plane will not be addressed in this paper.

Thus, the tetra-block determination as the minimal building unit for the structures approximated by tetrahedral chains, is proved by the supermapping of the minimal finite projective *PG(2,2)* geometry into the Riemann surface of Klein quartic, which is specified by the minimal hyperbolic Schwarz triangle

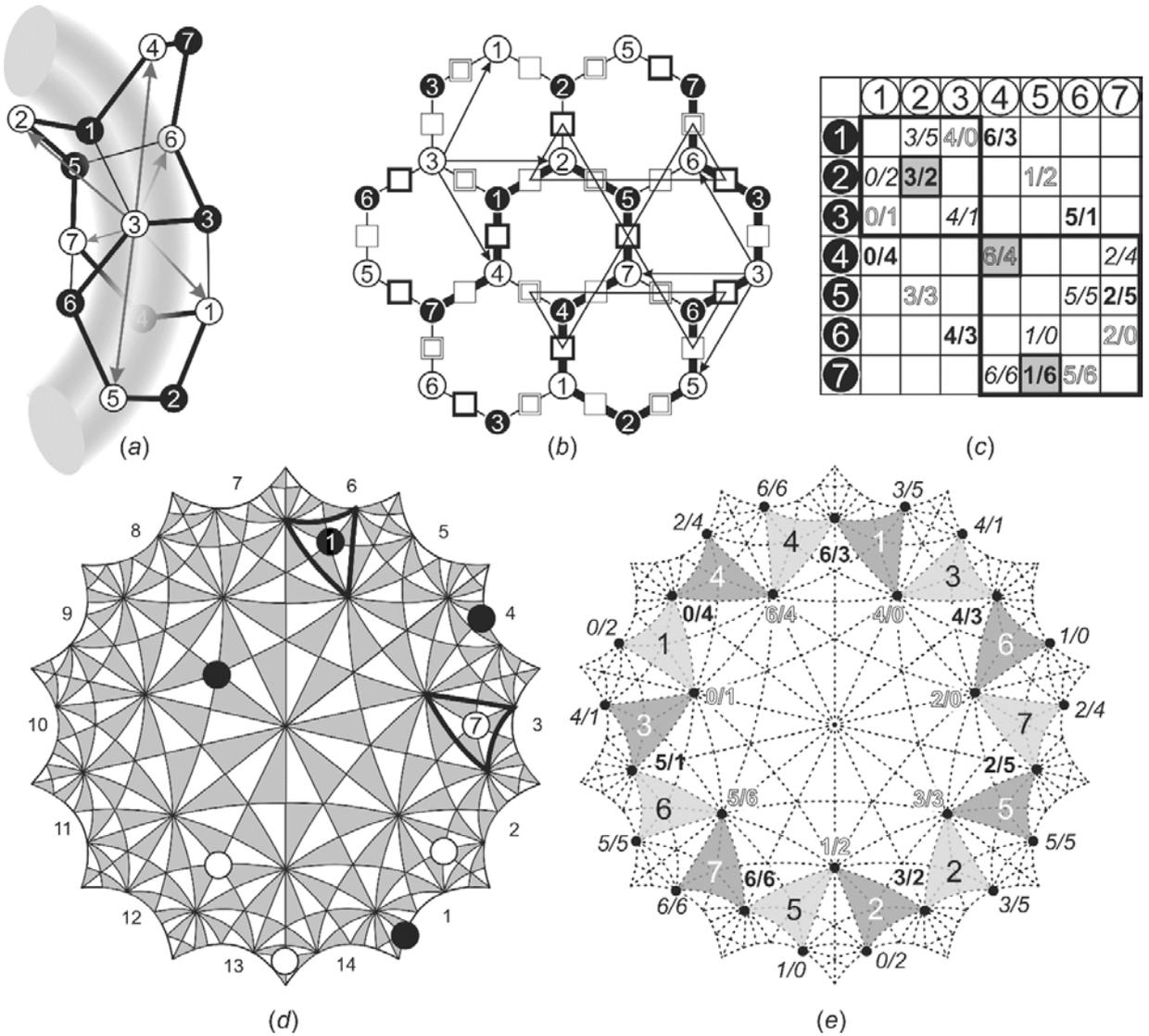

**Figure 3**. *a) compound 14-vertex tetra-block $\Delta_1^1$, where 6 hexacycles are partitioned by 18 edges. The union of white tetra-block $\Delta_1$ (fig 1.e.) and the black tetra-block $\Delta^1$ is realized by a double axis going through the midpoint of the edge 3-3'. The Hamiltonian graph $\Delta_1^1$ (shown heavy black line) coincides with the Hamiltonian graph in Figure 3b, only the vertex 3 is closest to all the other vertices of the tetra-block $\Delta_1$.*

*b) Map $\{6,3\}_{2,1}$ on torus is a regular 14 vertex bichromatic partitioning by 21 edges into 7 hexacycles. Equal vertices are identified. The Hamiltonian graph of the map is shown in heavy black line. As an example of adjacency of each point to the others the point 3 is chosen.*

*c) There is a one- to -one correspondence graph b) with the incidence table when the row number corresponds to a black dot, the column number to a white dot, and the incidence sign to a gray dot in the middle of an edge joining a white and a black vertex. An incidence sign is put into correspondence with a 4D vector exp $m(2\pi i/7)$, exp $n(2\pi i/7)$, represented as a fraction m/n. The vectors set to zero in order to define the tetra-block $\Delta_1$ are shown as gray squares.*

*d) Map $\{7,3\}_8$ on a sphere with 3 handles is a regular 56- vertex partition by 84 edges into 24 heptagons. Each heptagon is partitioned into 14 hyperbolic Schwarz triangles with angles $\pi/2$, $\pi/3$, $\pi/7$. The edges 1-6, 3-8, 5-10, 7-12, 9-14, 11-2, 13-4 are identified (Adapted from Figure1 in [23]).*

*e) Mapping PG(2,2) into Riemann sphere, constructed by modulating Klein's quartic by selecting 7 white and 7 black triangles out of 56 triangles (Adapted from Figure2 in [23]). Black triangle 1 and white 7 are selected in Figure a)*

## 5. Helix generated by tetra-block as a basis for structural model of the a-helix

A possibility of using the lattices $E_8$, $E_7$ and $E_6$ is related to the fact that the automorphism groups of densest lattice packings in dimensions ≤8 contain subgroups of small index generated by reflections (Coxeter groups). Existence of such subgroups allows for a use of local periodic groups for which integral representations over finite fields of algebraic numbers can be given [16]. In particular, the corresponding (projective) representations of the group $PSL(2,7)$ and its action on vector spaces lead to cyclotomic groups and the lattices related to them. Such are, for example, the Mordell-Weil (MW) lattices, based on using the set of rational (singular) points on the projective line or, as in the case in question here, on a minimal surface [27]. The lattice $MW(E^*_7)$ is a $7D$ sublattice embedded in $MW(E_8)$; its 1st coordination sphere contains 56 vectors corresponding to the vertices of $\{7,3\}_8$. It can be demonstrated that a tetra-block satisfying the relations (5), (6) is embedded in $MW(E^*_7)$. The tetra-block is part of the Coxeter helicoid (tetrahelix) – the union of tetrahedra by the axis $30/11$ given by the Petri polygon of the polytope $\{3,3,5\}$ [17]. The latter determines embedding of the tetra-block into the lattice $E_8$, which can be uniquely reproduced from the polytope $\{3,3,5\}$ [16]. Summing up the above,

$$(\mathbf{r}\{3,4,3\}\backslash\pm 8) \leftarrow Z[\zeta] \leftrightarrow D_4 \leftarrow E_8 \rightarrow MW(E^*_7) \rightarrow \{7,3\}_8 \rightarrow \{6,3\}_{2,1} \rightarrow \{6,3\}_{2,1}^{3(i)} \rightarrow \{3,6\}_{2,1}^{6(i)} \rightarrow \Delta_i, \quad (10)$$

the relations are obtained that determine the generation of the helix from the tetra-blocks $\Delta_i$. The arrow ↔ denotes an isomorphism between corresponding sets.

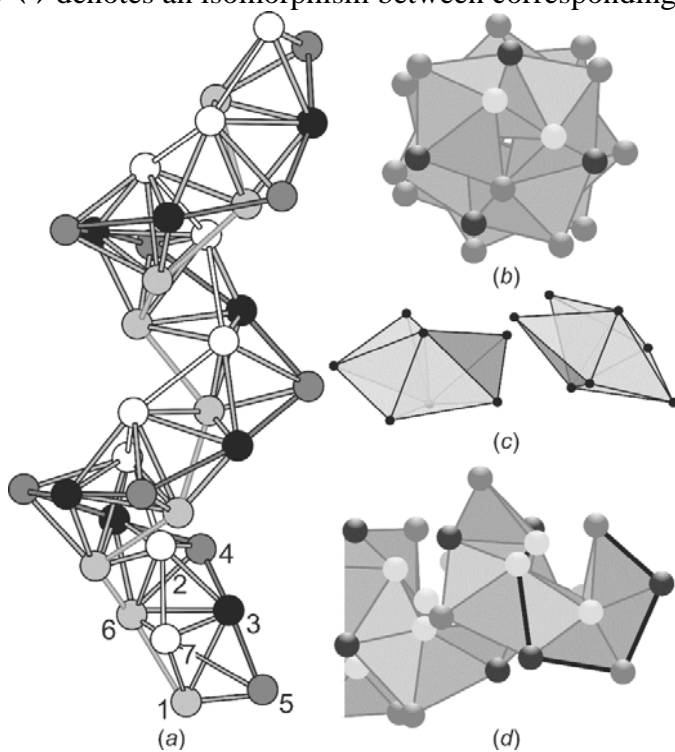

**Figure 4**. a) Helical joining of the tetra-blocks $\Delta_1$, in the face-to-face mode, tetra-block centers (vertices of the 3 type in Figure 2b) are colored black. White, black, grey and light-grey circles belong to 40/11 spirals. The same vertices are shown with the same color in all Figures.

b) View of helix in (a) along its axis. Light-grey vertices are invisible. Radius of the central empty channel is of 4% the spacing from the helical axis to black vertex (i.e. the radius of helix).

c) The joining of two flat $\Delta_3$ tetra-blocks with shaded common face into 11-vertex polyhedron.

d) The helix of $\Delta_3$ tetra-blocks generated by rule in Figure (c). Edges of the equatorial pentagon in the tetra-block $\Delta_3$ are shown in bold lines.

The (1,2,3,4) set of the tetrahedron vertices is subdivided onto four number triples (a face), and any number pair (an edge) belongs for two triples (faces) only. Therefore the tetrahedron $\Delta_0$ is defined by the *t-(v,k,λ)-scheme* of the block design, i.e. by the set of v elements subdivided onto blocks by k elements in each block. At such tiling any set of t elements presents into λ blocks exactly. The *t-(v,k,λ)*-scheme variant with $t=\lambda=2$, $v=1+k(k-1)/2$ is called a biplane. When $k=3, 4, 5$ biplanes constitute the special series of *2-(4, 3, 2), 2-(7, 4, 2), 2-(11, 5, 2)*. The defining construction *PG (2,2)* for a tetra-block is in correspondence with the combinatorial construction of the bi-plane *2-(7,4,2)* The *2-(7, 3, 1)* scheme is the complementary to the *2-(7,4,2)* scheme with the same group of automorphisms *PSL(2,7)* and the *PG(2,2)* is the geometrical interpretation of the *2-(7, 3, 1)* scheme. Generation of a helix by a tetra-block implies that each succeeding tetra-block attaches to the preceding one by face according to the same law. Within the approach of [28, 23] it can be established that the union by common face of two 7-vertex tetra-blocks into a 11-vertex figure is determined by the construction of biplane *2-(11,5,2)* with the group of automorphisms *PSL(2,11)*.

The joining of two tetra-blocks with a common face (Figure 2b) is possible with two fundamentally different variants: the identical (Figure 2c) or different (Figure 2d) chains of the *15/4* type are joined. Distinctly from an equal-edged tetra-block from the tetrahelix the tetrablock which is defined by the development in Figure 2a, has also shortened edges forming both the first and second chains of the *15/4* type (*1-3-2* and *4-6-7-5*). Hence, in the first joining variant 11-vertex fragment of the tetrahelix is obtained. The 40/11 spiral consisted from $\Delta_1$ tetra-blocks is defined by the second joining variant (Figure 4a, b). The delineation of diagonal quadrates 3 x 3 and 4 x 4 in the incidence table of the *PG (2,2)* (Figure 3c) corresponds to the delineation of short chains of the *15/4* type. The diagonal quadrates 3x3 and 4x4 reflect the presence of cyclic subgroups of $3^{rd}$ and $4^{th}$ orders in the *PSL(2,7)*.

In a general case it is possible to generate one and the same helix by various tetra-blocks, for example, $\Delta_1$ and $\Delta_3$ (Figure 4). In particular, upon rectifying in $E^3$ the Coxeter helix from the polytope {3,3,5}, the helix 40/11 appears consisting of isolated regular tetrahedra, winding around the empty cylinder, which can be partitioned by additional edges into a 4-helix of slightly deformed tetrahedra. Each of the isolated tetrahedra can be augmented either to $\Delta_1$ or to $\Delta_3$. Evidently, the additional symmetry *{(40/11)$\Delta_1$} = {(40/11)$\Delta_3$}* makes the helix generated in this way more stable. For example, while preserving the helix as a whole, small perturbations may lead to a transition from the partition of the helix into tetra-blocks $\Delta_1$ to its partition into tetra-blocks $\Delta_3$ (Figure 4a,d). Mutual transformation of the tetra-blocks $f\Delta_1 \leftrightarrow \Delta_3$ is realized by a Mobius transformation (4) determined by *PSL (2, 7)*. That is, a concretizing relation (6) is obtained defining generation of the helix from the tetra-blocks $\Delta_i$, $i=1, 3$ :

$$U \supset \Delta_\zeta \cup (\exp 2\pi i 11/40)\Delta_\zeta \rightarrow \Delta_1 \cup (40/11)\Delta_1 \leftrightarrow f\Delta_1 \cup f(40/11)\Delta_1 = \Delta_3 \cup (40/11)\Delta_3 , \qquad (11)$$

where $f(40/11)f^{-1}=(40/11)$, $f$ - automorphism of helix of tetra-blocks. The *(40/11)$\Delta_i$* denotes rotation of the starting tetra-block by the angle $\phi = 99°$ together with the shift by *11H/40* along the axis of a local cylinder of radius R, which transforms it into a tetra-block bordering the original one by shared face. The *f*-transformation is replacing each $\Delta_1$ by $\Delta_3$ and keeping intact the spiral consisting from tetra-blocks. The only change occurring in this *f*-transformation is the substitution of black vertices by the dark-grey vertices, and vice versa. Thus, the relations (10) – (11) determine a topologically stable helix with the axis 40/11 of tetra-blocks. It should be noted that empty central channel with the radius of *0.04 R* (Figure 4b) corresponds to the requirement of dense packing of helices [8].

For the case in question the centers of molecules (clusters) coinciding with the centers of tetra-blocks form a system of points on a spatial curve–the helix. The point $r_2=R\cos\phi$; $R\sin\phi$; $H(\phi/360)$ of the helix closest to the starting point $r_1=R;0;0$ may be obtained by rotation on the circle of radius $R$ by the angle $\phi$ along with the shift by $H(\phi/360)$, where $H$ is the pitch of the helix. The distance between these points is $L= (2R^2-2R^2\cos\phi+ (H(\phi/360))^2)^{1/2}$.

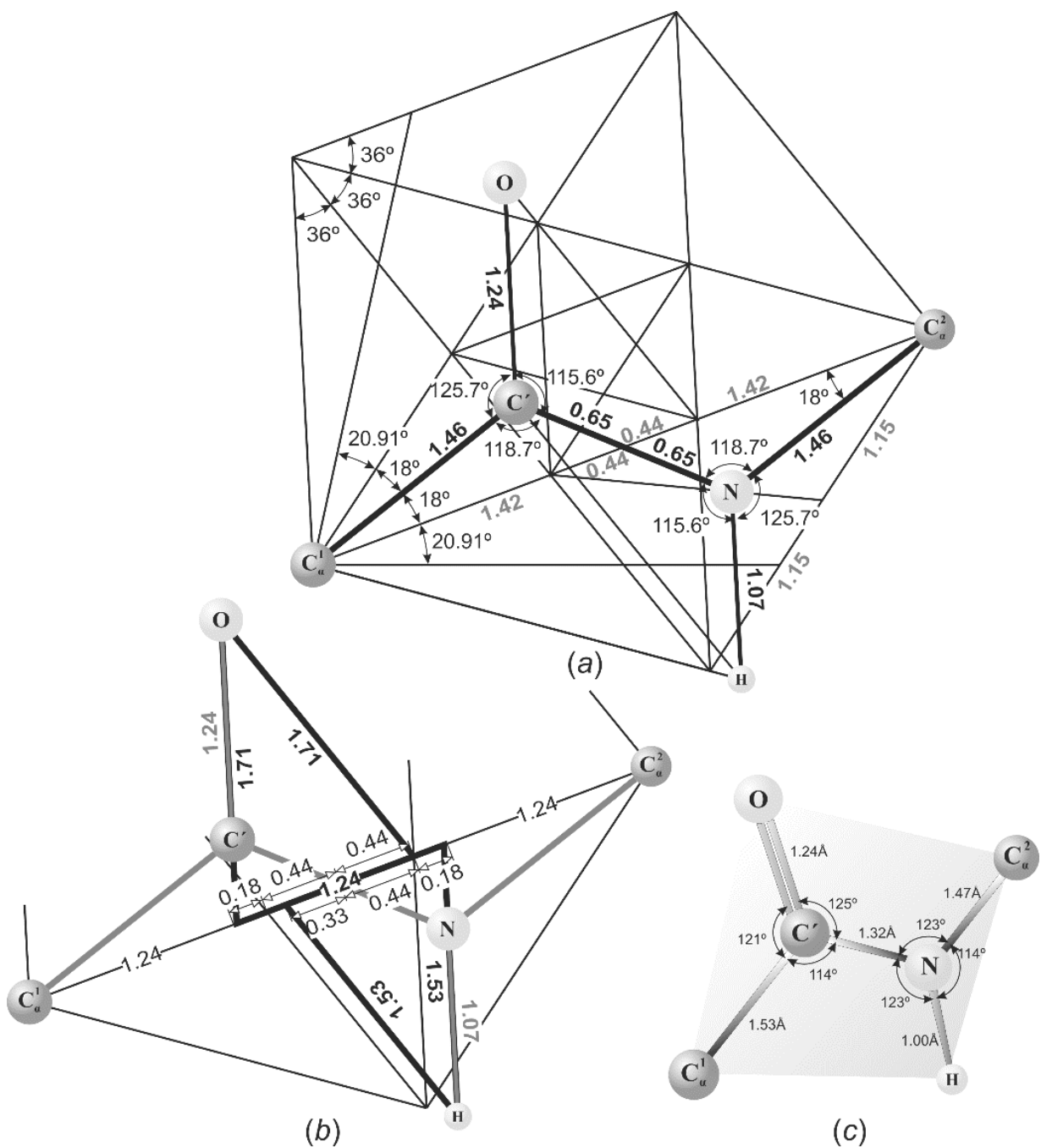

**Figure 5**. *a) Atoms of peptide plane of the α-helix are arranged in singular points of a regular pentagon - the equatorial section of the $\Delta_3$ tetra-block. Lengths of connecting edges are shown in Angstroms. $C^1_\alpha – C'$, $C' – N$ and $N – C^2_\alpha$ edges are symmetrical relative to the pentagon diagonal midpoint, C' and N- occupy the gravity centers (median intersections) of the golden triangles with edges 2.3.*

*b) An arrangement of O atom in the vertex of a golden triangle with the "silver basement" $1.06=2.3\,\tau^{-2}\,(1+\sqrt{2})$. Continuations of O-C' and H-N edges divide the pentagon diagonal into three parts, the C'-O edge is equal to one third of the diagonal. H atom is positioned in vertex of the golden triangle with the basement divided into ratio ≈1:2.*

*c) Experimental distances and angles in the peptide plane of the α-helix (Adapted from Figure 8 [29])*

According to experimental data for the protein α-helix: $R=2.3A$, $H_{exp.}=5.413A$, $H_{exp.}/R=2.3535$. With $\phi_{exp.}=99^0$ the spacing $L=3.8014A$, which is in accordance to the experimental value of $C_\alpha$-$C_\alpha$ spacing, and to the value of $H_{exp.}/2\pi R = tg\theta_{exp.}$, with the pitch angle for α-helix $\theta_{exp.}\approx 20.535^0$. With values of $H/R=2.4$, $\phi=99^0$ and $R=2.3$ A satisfying to (2), the spacing between tetra-blocks $\Delta_1$ centers (black circles in Figure 4a) $L=1.658 R$ is equal to $3.8134$ A. The pitch angle $\theta_{cr.}\approx 20.906^0$ is close to the pitch angle $\theta_{exp.}=20.535^0$ of the α-helix, therefore the helix, formed by $C_\alpha$ atoms from α-helix, could be approximated by the topologically stable helix constituted by the tetra-block $\Delta_1$ centers (11) with $R=2.3A$, $H=5.52A$, $\phi=99^0$, $H/R=2\pi/\tau^2$.

All starting $C_\alpha$, $C'$, $N$, $O$ and $H$ atoms are arranged into the same peptide plane. The symmetry plane presents in the flat $\Delta_3$ tetra-block only, therefore to associate atoms in the peptide plane with the certain symmetry distinguished points of tetra-block, it is necessary to transit by the relation (11) to the helix of $\Delta_3$ tetra-blocks. In this case two neighboring $C^1_\alpha$ and $C^2_\alpha$ atoms are two vertices of the pentagon diagonal in the equatorial section of the $\Delta_3$ tetra-block. For simplification this pentagon is regarded as the regular one, therefore with the unit pentagon edge the diagonal is divided onto three parts as: $\tau^{-1}$, $\tau^{-2}$ and $\tau^{-1}$. In the helix (Figure 4) the $\Delta_3$ tetra-block consists of two pairs of tetrahedra from two neighboring $\Delta_1$ tetra-blocks, and at that the common plane in the pair of $\Delta_1$ intersects the equatorial plane of the $\Delta_3$ tetrahedron along the bisectrix of the pentagon angle. Two neighboring developments of the $\Delta_1$ tetra-blocks inside the *U-strip* (Figure 2a) are joined with the local two-fold axis, therefore the required points of the polypeptide chain must be defined by the pentagon tiling which is symmetrical relative to the two-fold axis. That two-fold axis is orthogonal $C^1_\alpha$-$C^2_\alpha$ diagonal passing through its midpoint.

Construction of the Pythagorean pentagram is effected by inscribing into the pentagon, this operation permits to delineate isosceles golden triangles with the $36^0$ angle between unit edges which are satisfying noted conditions. In case $C'$ and N atoms are positioned in the gravity centers of such triangles one obtains a symmetrized fragment of the polypeptide chain. When the pentagon edge is equal to the spiral radius $R=2.3A$, the $C^2_\alpha$-$N$, $N$-$C'$ and $C^1_\alpha$-$C^2_\alpha$ spacing of such fragment are coinciding with the experimental values, and the $N$-$C^2_\alpha$ spacing is equal to the $C^1_\alpha$-$C'$ spacing (Figure 5a,c). In the vertex of the golden triangle with the «silver basement» O atom is positioned. This basement consists of $\tau^{-2}/2$, $\tau^{-2}/2$ and $(\tau/3-\tau^{-2})/2$ fractions of the pentagon diagonal, and ratio $\tau^{-2}/2 : (\tau/3-\tau^{-2})/2 = 3\tau/2\approx(1+\sqrt{2})$ defines the silver sections appearing by taking into account the relationship (6) between tetra-block with the $Z[\zeta]$ lattice. The $C'$-$O$ spacing is equal to $\tau/3$, atom H is positioned in the vertex of the golden triangle with the basement divided into $1:2$ relation. The $N$-$H$ spacing is equal to $(1+\sqrt{2})\tau^{-2}/2$ (Figure 5b). The experimental value of the $N$-$C^2_\alpha$ spacing is less than the $C^1_\alpha$-$C'$ spacing, it could be associated with the insignificant declining from the regularity of the pentagon in the equatorial section of the $\Delta_3$ tetra-block. In this fashion atomic positions in the peptide plane of the α–helix is defined by singular points of Pythagorean pentagram in the equatorial section of the $\Delta_3$ tetra-block (Figure 5).

Geometric factors in Figure 5 show that the α-helix stability conditions are determined by the juxtaposition of electronic bond lengths with the geometric parameters of the tetra-block viz its unit edge. It could be shown that the symmetries delineating these special points are determined by the *PSL(2,11)* group. In accordance to (2) the angle between the pentagon diagonal $C^1_\alpha$-$C^2_\alpha$ and horizontal is equal to $20.91^0$. By taking the angle of $20.91^0$ with the symmetry relative to the $C^1_\alpha$-$C'$ line, one obtains the angle of $18^0 + 18^0+20.91^0=56.91^0$ between line $C^1_\alpha$-$C^5_\alpha$ (Figure 2a) and horizontal. This angle characterizes the 40/11 axis. It is remarkable that the intrinsic to the Pythagorean pentagram the angle of $18^0$ between $C^1_\alpha$-$C^2_\alpha$ edge and horizontal coincides with pitch angle of the spiral with the maximal packing density [8].

## 6. Conclusion

It is established in the present work that topologically stable in $E^3$ structures are determined by a triangulated discrete surface [14] corresponding to the bifurcation point of a catenoid and given by a Weierstrass representation by the minimal surface $M_0$ with zero instability index. A chain of constructions of algebraic geometry and topology makes possible a transition from $M_0$ to a topologically stable helix with the pitch to radius ratio equal to $2\pi/\tau^2$. In view of the local cylindrical nature of $M_0$ and its definition as a surface common to both the catenoid and the helicoid, the radius of a topologically stable helix of spheres must be close to the radius of the sphere and there must exist a small sized empty central channel. The 80 – vertex "polytope" determined by $M_0$, as well as the Mordell-Weil $E_7$ – like lattice embedded in $E_8$ allow for selection of the local axis of helical rotation $40/11 = 40\exp(-2\pi/\tau^2)$. This parametric axis provides for a distribution of 40 vertices into 11 turns of the helix, and is simultaneously a period (parameter) of the local lattice, like a translation is the period for a crystal. Homogeneity of the helix is ensured by partitioning the 40 vertices into 4 cycles of 11 vertices, each of which borders its neighbor on a common vertex.

While every polyhedron may be partitioned into tetrahedra joined by their faces, in $E^3$ regular tetrahedra can be packed only in chains. The considered helical biostructures belong to a broad class of structures determined by chains of regular tetrahedra. For such structures it has been possible to define a mathematically determined building unit – a tetra-block, realized in 3 variants. The universality and topological stability of the tetra-block is caused by the possibility to embed in $E^3$ non-Euclidean (but locally Euclidean) substructures determined by $M_0$ with groups of fractional (non-rigid) linear transformations. The minimality of the 7 – vertex tetra-block as a construction entity follows from its unique defining mapping from the Klein's quartic (specified by minimal hyperbolic Schwartz triangle) into the minimal finite projective geometry $PG(2,2)$. The helix generated by the tetra-block determines the parameters of an ideal, mathematical $\alpha_0$ – helix when putting into correspondence to the atom $C_\alpha$ a center of the tetra-block $\varDelta_1$. The helix generated by the tetra-block $\varDelta_1$ is also generated by the tetra-block $\varDelta_3$, which is transformed into $\varDelta_1$ by a transformation of the Mobius geometry. Partitioning the helix into both $\Delta_1$ and $\Delta_3$ makes it still more stable and allows to determine the positions of the $N$, $C_\alpha$, $C'$, $O$, $H$ in the equatorial planes of $\varDelta_3$ - peptide planes of the $\alpha_0$ – helix.

Experimentally determined structural parameters of the $\alpha$–helix represent a realization (in the approximation of a local cylindrical parametric axis (for the hyperbolic circle) of the parameters of the $\alpha_0$–helix determined by the relations (1),(2),(7),(8),(11). To confirm this let us put into correspondence the main structural parameters of the $\alpha_0$–helix and the $\alpha$–helix:

1) The ratio of the helical pitch to radius $H/R=2.4$ and $H/R=2.35$; the pitch angle $20,9^0$ and $20.5^0$, the distance $C_\alpha – C_\alpha$ $3.81A$ $(3.72A)$ and $3.80A$.
2) The axis of helical rotation 40/11 (rotation by $99^0$) and 36/10 (rotation by $100^0$)
3) The relation $i \to i+4$ determining the positions of hydrogen bonds in the $\alpha$-helix is a realization of the relation $(40/11)^4 = 10_1$.
4) Experimentally observed average length of the *α*-helix of 11 residues is determined by the relation $(40/11)^{10} = 4_1$ giving a partition of 40 vertices into four 11– vertex cycles with one common vertex.
5) Atomic positions in the peptide plane of the $\alpha$–helix are determined by special positions in equatorial section of the tetra-block $\varDelta_3$.
6) Approximately coinciding edge of tetrahedron of tetra-block and radius $R = 2.3A$ of $\alpha_0$–helix.
7) Existence of the central empty channel with the radius of $0.04R$ in the $\alpha_0$-spiral assembled from tetra-blocks.

The crystal structure can be considered as the polyhedra orbit of the space group when the starting polyhedron is decorated by atoms. Similarly, $\alpha$–helix is the tetra-blocks orbit of the 40/11 axis defined by the Mordell-Weil $E_7$ lattice in case the starting block is decorated by atoms.

The local axes considered in this paper belong to the class of 35 parametric axes determined by the lattice $E_8$ [20]. This class also contains the axes 10/3=3,(3); 30/13=2,(307692). According to table 5.1 [1] the number of residues per turn equals: in the helix of collagen fibers – 3.3, in a twisted β – structure – 2,3. Thus, for the basic polypeptide chains the lattice $E_8$ *a priori* determines local parametric axes like crystal lattice determines possible screw axes of crystals. However, not all of them are topologically stable, which determines, for instance, the instability of the $3_{10}$–helix as well as hypothetical character of the π– helix. Table puts into correspondence the theoretical (ours and [8]) and experimental [1] data for the helices of polypeptide chains.

**Table.** Experimental [1] and theoretical (ours and [8]) parameters of helices of polypeptide chains.

| Helices of polypeptide chains. | Residues per turn | Local axis of the mathematical helix | Helical pitch H (A) | Helical radius R (A) | Helical pitch angle θ($^0$) | | | Volume fraction for a packed helix |
|---|---|---|---|---|---|---|---|---|
| table 5.1 [1] | [1] | relation (**7**) | [1] | [1] | (2) | [1] | [8] | [8] |
| β-structure | 2.3 | 30/13=2,(307692) | 7.59 | 1.0 | | | | |
| $3_{10}$- helix | 3 | 30/10=3,(0) | 6 | 1.9 | | 26.6 | | 0.690 |
| collagen helix | 3.3 | 40/12=10/3=3,(3) | 9.67 | 1,6 | | | | |
| **α- helix** | **3.6** | **40/11=3,(63) = =40 exp(-5.5/2.3)** | **5.4** | **2.3** | **20.9** | **20.5** | **18.1** | **0.781** |
| π- helix | 4.3 | 30/7= 4,(285714) | 4.73 | 2.8 | | 15.0 | | 0.777 |

At the present time the structural classification of proteins is primarily based on bioinformatics using capabilities of computer enumeration and allowing direct comparison between proteins [2]. The formalism employed in this paper (its detailed mathematical exposition is given in [30,31]) allows for discovery, in advance of any real or computer experiments, of symmetry–related laws of structure of some classes of biopolymers using possibilities of a priori selection of "topologically stable" structures. Firstly, such an approach is relevant to systematization of structures of the principal elements of protein structures, which still a subject of discussion and is, in fact, only beginning to become formalized. Although this work considers just the α-helix, the general approach being developed allows us to obtain results also on symmetry foundations of stability of *A, B, Z – DNA*. It is evident that possible but not yet experimentally discovered symmetries might point to experimentally missed (or still lacking adequate explanation) solutions, for instance, for the structures and transformations of the corresponding forms of DNA.

**References**
1. Shulz G.E., Schirmer R.H. Principles of Protein Structure. NewYork: Springer-Verlag. 1979.
2. Finkelstein A.V., Ptitsyn, O.B. Protein Physics. Amsterdam - Boston - London - New York : Academic Press. 2002.
3. Sadoc F. J., Rivier N. Boerdijk-Coxeter helix and biological helices // Eur. Phys.J. 1999. B12, P.309-318.
4. Sadoc F J. Helices and helix packings derived from the {3; 3; 5} polytope // Eur. Phys. J. 2001. E5, P.575-582
5. Nelson, D. L., Cox, M.M. Lehninger Principles of Biochemistry. 4th ed., New York: W.H. Freeman & Co. 2004
6. Pauling, L., Corey R. B., Branson H. R. The structure of proteins: two hydrogen-bonded helical configurations of the polypeptide chain. // Proc Natl Acad. Sci. USA 1951. V.37, P.205-211.

7. Banavar J.R., Maritan A.Physics of proteins. Ann. Rev. Biophys. Biomol. Struct. 2007, V.36, P.261-280
8. Olsen K., Bohr J. The generic geometry of helices and their close-packed structures // Theor. Chem. Acc. 2010, V.125, P.207-2015.
9. Lu C. H, Huang S. W, Lai Y. L, Lin C P, Shih C H, Huang C C, Hsu W L, Hwang J K. On the relationship between the protein structure and protein dynamics. //Proteins.2008 ;V 72, P.625-634.
10. Yee D. P, Chan H. S, Havel T. F, Dill K.A. Does compactness induce secondary structure in proteins? A study of poly-alanine chains computed by distance geometry// J Mol Biol. 1994 ; V.241 P.557-573.
11. Samoylovich M.I., Talis A.L. Symmetry of helicoidal biopolymers in the framework of algebraic geometry: α- helix and DNA structures.// Acta Cryst. 2014. A.70, P.186-198.
12. Fomenko A.T., Tuzhilin A.A. Translations of Mathematical Monographs,Vol.93, Elements of the Geometry and Topology of Minimal Surfaces in Three-Dimensional Space. Providence: American Mathematical Society. 1992.
13. Busemann H., Kelly P. Projective geometry and projective metrics. New York Academic press inc. publishers. 1953
14. Lam W. Y., Pinkall U. Isothermic Triangulated Surfaces // arXiv:1501.02587v2 , 2015.
15. Bobenko A, Pinkall U. Discrete Isothermic Surfaces //J. reine angew. Math. 1996, V.475, P.187-208
16. Conway J.H., Sloane, N.J.A. Sphere Packing, Lattice and Groups. New -York : Springer-Verlag,1988.
17. Coxeter H.S.M. Regular polytopes. New - York : Dauer, 1973
18. G. de B.Robinson. On the orthogonal groups in four dimensions **//** Proc. Camb. Phil. Soc. 1931, V.27, P.37-48.
19. Dubrovin B.L., Novikov S.P., Fomenko, A.T. Modern-Day Geometry Moscow: Éditorial URSS. 2001.
20. Samoylovich M. I., Talis A.L. Gosset helicoids: II. Second coordination sphere of eight - dimensional lattice $E_8$ and ordered noncrystalline tetravalent structures// Crystallography Reports. 2009, V.54, P.1117 - 1127.
21. Coxeter, H.S.M. Introduction in geometry. New-York- London: Jonh Wiley & Sons.1961.
22 Babiker H., Janeczko S. Combinatorial cycles of tetrahedral chains/ IM PAN Preprint 741, 2012
23. D.Martín P., & Singerman D (). The geometry behind Galois' final theorem// Eur.J. of Combinatorics, 2012, V.33, P.1619-1630.
24. Lord E. A., Mackay A.L., Ranganatan S. New geometry for new materials/ Cambridge university press. 2006.
25. Elkies N. The Klein Quartic in Number Theory / The Eightfold Way. MSRI Publications 1998, V. 35, P.51-100.
26. Karteszi F. Introduction to Finite Geometries. Akademiai Kiado, Budapest, 1976.
27. Shioda T. Theory of Mordell-Weil Lattices / Proceedings of the International Congress of Mathematicians, Kyoto, Japan, 1990, P. 473-489.
28. Brown E. The fabulous (11, 5, 2) biplane. // Math.Mag. 2004, V.77, 67–100.
29. Fraser R. M. A Tale of Two Helices. A study of alpha helix pair conformations in three-dimensional space. Kingston, Ontario, Canada 2006.
30 . Samoylovich M.I., & Talis A.L (2012). ArXiv:1211.3686; ArXiv:1211.6560,
31. Samoylovich, M.I., & Talis, A.L (2013) ArXiv:1303.4228; ArXiv 1312.7107